\documentclass[journal,onecolumn]{IEEEtran}

\usepackage[utf8]{inputenc}
\usepackage{multirow}
\usepackage{graphicx}
\usepackage{indentfirst}
\usepackage[small]{caption}

\ifCLASSINFOpdf
\else
\fi

\begin{document}
%
\title{Super-Resolution Image Reconstruction Based on Self-Calibrated Convolutional GAN}
%
%
%

\author{Yibo~Guo,
        Haidi~Wang, Yiming~Fan, shunyao ~Li, 
        and~Mingliang ~Xu,
\thanks{Y. Guo is a teacher at the School of Information Engineering, Zhengzhou University}
\thanks{H. Wang, Y. Fan, S. Li and M. Xu are with Zhengzhou University.}
}

%
%

\markboth{Journal of \LaTeX\ Class Files,~Vol.~14, No.~8, August~2015}%
{Shell \MakeLowercase{\textit{et al.}}: Bare Demo of IEEEtran.cls for IEEE Journals}
%



\maketitle

\begin{abstract}
With the effective application of deep learning in computer vision, breakthroughs have been made in the research of super-resolution images reconstruction. However, many researches have pointed out that the insufficiency of the neural network extraction on image features may bring the deteriorating of newly reconstructed image. On the other hand, the generated pictures are sometimes too artificial because of over-smoothing. In order to solve the above problems, we propose a novel self-calibrated convolutional generative adversarial networks. The generator consists of feature extraction and image reconstruction. Feature extraction uses self-calibrated convolutions, which contains four portions, and each portion has specific functions. It can not only expand the range of receptive fields, but also obtain long-range spatial and inter-channel dependencies. Then image reconstruction is performed, and finally a super-resolution image is reconstructed. We have conducted thorough experiments on different datasets including set5, set14 and BSD100 under the SSIM evaluation method. The experimental results prove the effectiveness of the proposed network.
\end{abstract}


\IEEEpeerreviewmaketitle

\section{Introduction}
%
%
%
%
Single image super-resolution reconstruction (SISRR) is an important research direction in the field of computer vision. Super-resolution reconstruction aims to reconstruct the corresponding high-resolution image from the observed low-resolution image to obtain richer image detail information, which has been utilized in medical imaging\cite{2019Human,2020Unbalanced}, monitoring equipment\cite{2018Landslide} and other fields\cite{2018Finding}.

The main task of SISRR is to determine the mapping function between low-resolution image and high-resolution image, and reconstruct the high-resolution image corresponding to the low-resolution pictures. With the development of convolutional neural network in the image super-resolution, SISRR techniques have been developed rapidly. Among them, SRCNN\cite{Chao2014Learning} is the first technique to use convolutional neural network to process super-resolution image reconstruction, which solved the ill-posed problem caused by the traditional method of learning mapping function between low resolution and high resolution.  With the gradual rise of Deep Learning, the newly designed structures based on deep networks have achieved good results by extracting much more features from the original pictures. However, deep networks require relatively long training time while many detailed information which has been extracted by the complex deep network cannot be made full use of.


In this paper, we proposes a self-calibrated convolutional network structure based on a generative adversarial networks in order to make full use of the extracted features. The generator structure consists of two parts: feature extraction and image reconstruction. Feature extraction introduces a self-calibrated convolutional neural network\cite{Liu2020Improving} to fully extracts the features of the low resolution image. Firstly, a multi-scale method is adopted to the feature fusion module. The fusion results are the input of the image reconstruction part, while the reconstruction result are the output. Compared with other methods such as spatial pooling and attention\cite{Irwan2020Attention,Hu2018Gather,linsley2018learning}, which are utilizing the convolution operation for extracting long-range dependencies, the method in our paper divides the learnable convolution kernel into four parts. By doing this, the new kernel can not only adaptively encode the context information in the long-range region as well as the spatial position feature of the pixel in the image efficiently, but also obtains the dependence between the channels in each spatial position. This method can be simply embedded in a common convolutional neural network\cite{1988Neocognitron} without adding any hyper-parameters.


In addition, super-resolution image reconstruction usually uses mean squared error(MSE)\cite{2016Image} as the loss function.  However, MSE is highly sensitive to large errors and has limited processing at the pixel level of the image, which may make the reconstructed image be over-smoothing. Therefore, some scholars proposed to use structural similarity index(SSIM)\cite{Zhou2004Image} instead of MSE as the loss function. SSIM mainly focuses on the brightness, contrast, and structure, etc, and is mainly used to evaluate the similarity of two image. This paper adopts an adaptive robust loss\cite{Barron2019A}, which learns hyper-parameters independently, and reduces the workload of manual tuning. The function form is not only limited to MSE, but also includes L1 loss, L2 loss, and various loss functions Combines, which is conducive to network training with good robustness.

In order to alleviate the above mentioned problems, this article mainly made the following contributions:

\begin{itemize}
\item Improved the generator structure in SRGAN\cite{Ledig2016Photo}. The generator network includes two components: feature extraction and image reconstruction, which simplifies the network complexity.
\item The self-calibrated convolutional network is first time to be introduced in the feature extraction part of the generator, and it can be used to reconstruct the super-resolution task to fully extract the image features and make full use of the detailed information.
\end{itemize}

\section{Related Work}
\subsection{Super-resolution image reconstruction based on convolutional neural network}

The method in this paper mainly uses convolutional neural networks, including shallow neural networks and deep neural networks, to learn the mapping relationship between low-resolution image and high-resolution image. SRCNN\cite{Chao2014Learning} is the first to use CNN as the mapping function. In order to achieve the image reconstruction, the bicubic interpolation is performed  before the feature extraction. This expands low-resolution images at the beginning, which increases the cost of network calculation. FSRCNN\cite{Chao2016Accelerating} is an improved version of SRCNN. The activation function Relu is replaced by Relu with parameters, which is more conducive to network training. Low-resolution images are directly input into the network, and the image reconstruction is performed at the end of the network. This process solves the problems of SRCNN of which the computational complexity is overwhelming. However, it is not yet able to fully extract the features of the image.

With the development of deep learning, the number of network layers of the newly emerged structures continues to increase. The lapSRN\cite{Lai2017Deep} utilizes the Laplacian pyramid structure to perform deconvolution operations on each layer of the pyramid in order to achieve up-sampling. This method exploits the nature of Laplacian pyramid to predict the sub-band step by step, and are applied with the cascaded convolution operation to extract the features. DRCN\cite{2016Deeply} is the first technique uses the recurrent convolution network for super-resolution image reconstruction. The idea is to determine the appropriate number of recurrent convolution layers to prevent gradient disappearance and gradient explosion in order to extract the high frequency of the image information. VDSR\cite{2016Accurate} is based on an improved 20-layers VGG network, of which taking into account the network convergence slowly caused by network deepening. A cascaded convolution kernel is adopted to learn the high-frequency residual information between low-resolution and high-resolution images. This method requires to be combined other training techniques, such as the gradient clipping strategies. EDSR\cite{Lim2017Enhanced} improves the residual network structure and increases the network depth while removes the unnecessary modules in the network. The deep network solves the problem of instability in training with some certain advantages, but it is still suffered by the increase of the occupation of resources. EDSR is a single-scale extraction of high-frequency information, while VDSR is a multi-scale extraction of high-frequency information. Both of them needed bicubic interpolation to process the input.

In summary, although the reconstructed super-resolution image is relatively fuzzy, the shallow network is lack of the ability of fully extracting the features. On the other hand, the Relu with parameters and the deconvolution method have been widely applied with the subsequent super-resolution image reconstruction in researches. Even if the deep convolutional network can extract more information, it cannot make full use of the extracted feature information. In order to solve the problems of complicated calculations caused by deepening of the network, Wang \emph{et al.} \cite{He2020ODE} proposed to use the ordinary differential equations to guide network design. Compared with the above works, the self-calibrated network proposed in our paper does not require additional parameters, and the image features can be effectively extracted and fully utilized.  Our proposed method can achieve the same effect as a complex networks, and can be embedded into different tasks.

\subsection{Super-resolution image reconstruction based on generative adversarial network}

Generative Adversarial Network (GAN)\cite{Goodfellow2014Generative} is proposed by Goodfellow in 2014. It includes a generator and a discriminator. The generator is used to generate diverse samples, and the discriminator was essentially composed by a two-classifier structure in order to identify image samples whether the generated images are real. The generator and the discriminator perform a min-max game, and finally the generator generates samples with diversity and authenticity. DCGAN\cite{2015Unsupervised} is a popular method which introduces the Relu with paramters and deconvolution methods into the GAN structure. MAD-GAN\cite{Ghosh2017Multi} proposed a multi-generator and single discriminator structure, and the parameters are shared among multiple generators. Relevantly, the discriminator also made corresponding changes to match the generator changes. The model is mainly designed to prevent to occuring the problem of mode collapse. DGAN\cite{Zareapoor2019Diverse} used MAD-GAN for super-resolution image reconstruction, and obtained best results. SRGAN\cite{Ledig2016Photo} is based on a deep residual network module to obtain the context information of the image. The utilization of the jump connections increases the complexity of the network. The multi-scale generative adversarial network\cite{Liu2020Image} is proposed as a improvement method based on SRGAN. Compared with SRGAN, this method has a deeper network structure with more numbers of layers but unsatisfied results.

\subsection{Loss function}

In the network training process, choosing a suitable loss function can help the network converge quickly and better learn the distribution characteristics of the data. The L1 loss was the sum of the absolute value of the difference between the predicted value and the target value in the regression task. The loss function The disadvantage was that the position of the center point was not derivable. In the training of the deep neural network, the objective function took the derivative and then propagated back, but this function was not good for the calculation and solution. MSE\cite{2016Image} was to calculate the sum of squares between predicted value and the target value. The use of the MSE loss function in the super-resolution image reconstruction task caused the reconstructed image to be too smooth, and the human visual effect was poor. So some scholars proposed to used SSIM\cite{Zhou2004Image} as the loss function instead of MSE , SSIM is usually used as an evaluation method in super-resolution reconstruction tasks to evaluate the similarity of reconstructed image in brightness, contrast, structure, etc. It obtained certain effects. SRGAN\cite{Ledig2016Photo} proposed a perceptual loss function based GAN structure. The perceptual loss function included content loss and adversarial loss. The content loss included the use of MSE to calculate the similarity between the reconstructed image and the real image and the similarity between the high-level features of the image by VGG network. MSE accounted for the dominant content loss. Therefore, the above-mentioned reconstructed image still appeared too smooth.

\section{Method}

\subsection{Problem definition}

The main purpose of our research is to reconstruct a single low-resolution image into a super-resolution image using a generative adversarial network, which is defined as follows: the real high-resolution image ${I^{HR}}$ is expressed as rW*rH*C, and C is the number of image channels; the input image is low-resolution image ${I^{HR}}$ expressed as W*H*C, which is obtained by down-sampling the real high-resolution image ${I^{HR}}$; and the reconstructed super-resolution image ${I^{SR}}$ is expressed as rW*rH*C. Using the generative adversarial network to learn its mapping function, the reconstruction result expresses as ${I^{SR}} = G({I^{LR}})$. The GAN objective function is modified as:

\begin{center}
$\mathop {\min }\limits_G \mathop {\max }\limits_D V(D,G) = {E_{{I^{HR}} \sim {p_{data}}({I^{HR}})}}\log ({D_{{\theta _d}}}({I^{HR}})) + {E_{{I^{LR}} \sim {p_{data}}({I^{LR}})}}\log (1 - {D_{{\theta _d}}}({G_{{\theta _g}}}({I^{LR}})))$
\end{center}

Among them, D is the discriminator, G is the generator.
\subsection{Method structure}

A 3*3 convolution kernel is usually chosen for feature extraction in computer vision applications. In order to extract more detailed information from the picture, it is possible to choose a 5*5 convolution kernel instead of a 3*3 convolution kernel to increase the receptive field, which increases the network parameters. Also, some researchers have proposed that two 3*3 convolution kernels are equal to a 5*5 con volution kernel, while the parameters are less than 5*5 convolution kernels. In MSRN\cite{Li2018Multi}, the authors bulid an MSRN module with the 3*3 convolution kernels and the 5*5 convolution kernel. The two convolution operations in the module extract image features separately, and the extracted features are merged at different scales. This method can effectively extract image features, and the quality of reconstructed image is satisfactory. However, multiple convolution operations increase the network parameters. The generator structure in our method uses a self-calibrated convolution network. It uses a 3*3 convolution kernel to fully extract image features, and uses feature fusion to make full use of the extracted features.

Our propose is to reconstruct the overall structure of the low-resolution image method as shown in Figure 1. The input of the traditional generative adversarial network are Gaussian noises. Since our paper uses the generative adversarial network to learn the mapping between low-resolution images and high-resolution images, the input of the generator is changed to a low-resolution image. The generator network does not generate new diversity samples, but reconstructs super-resolution image samples. The input of the discriminator are reconstructed super-resolution images and real images. The output of discriminator is the discriminant results. After iterative training, the generator and discriminator in the model reach the Nash equilibrium, and the reconstructed sample is indistinguishable from the real sample.
\begin{figure}[t]
\centering
\includegraphics[scale=0.6]{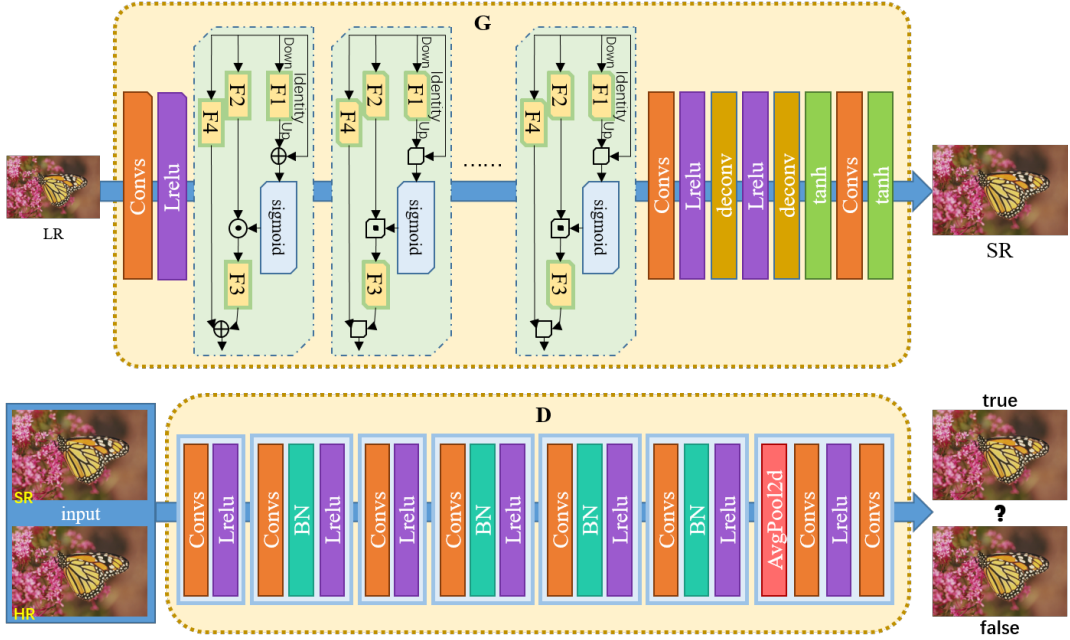}
\caption{The architecture of our proposed self-calibrated GAN.}
\end{figure}

\subsubsection{Generator structure}

The GAN structure in our paper is improved on the basis of the structure of SRGAN\cite{Ledig2016Photo}. SRGAN uses residual network to extract image features. However, its complex network structure does not utilizes the feature sufficiently. The generator network in our method includes two operations: feature extraction and image reconstruction operation. With simplified complexity of the network, it can still extract sufficient features and make full use of them. The feature extraction operation uses a self-calibrated convolutional network\cite{Liu2020Improving}, which is a set of convolution operations divided into four portions. Each portion of the convolution setting is similar to ordinary convolution but has specific functions. The detailed structure diagram is shown in Figure 2. The process is divided into two modules: The first module contains three portions, which are \textcircled{1}, \textcircled{2} and \textcircled{3} in figure 2 respectively. The first portion is down-sampling and up-sampling to extract the image features. The second portion directly extracts the image features. After that, the results of the two portions are merged. In the third portion, spatial information of different scales can be obtained and the range of receptive fields are enlarged. The second module contains the fourth portion, corresponding to \textcircled{4} in figure 2. The operation is as follows: The fourth portion directly extracts the features, and its function retains the original spatial context information. Finally, two modules are connected together. In this method, we embed multiple operations in the generator, of which the purpose is to fully extract image features. Putting the reconstruction part after the feature extraction can save training time and resource utilization. The reconstruction part adopts deconvolution operation, and the form is more concise.

\begin{figure}[t]
\centering
\includegraphics[scale=0.6]{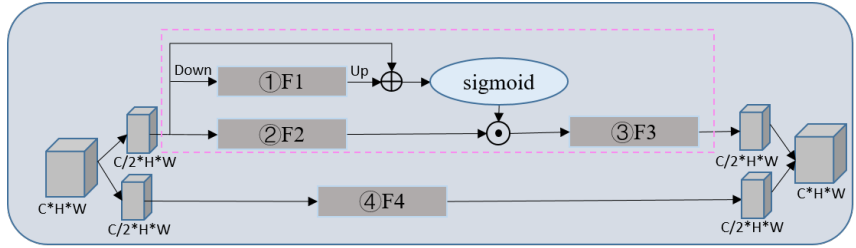}
\caption{The architecture of detailed generator}
\end{figure}

Therefore, the combined calculation formula of self-calibrated convolution is:

\begin{center}
${\rm{mid}} = {f_3}({f_2}(x) \times sigmoid(x + up({f_1}(x))))$
\end{center}

Among them, ${f_1}$, ${f_2}$ and ${f_3}$ represent the convolution operations shown in \textcircled{1}, \textcircled{2} and \textcircled{3} in figure 2, respectively, and $up$ represent the up-sampling operation.

The final generator result is expressed as:

\begin{center}
${I^{SR}} = G({I^{LR}})$
\end{center}

The adversarial loss of the generator is calculated as follows:

\begin{center}
$\mathop {\min }\limits_G  = {E_{{I^{LR}} \sim {p_{data}}({I^{LR}})}}\log (1 - {D_{{\theta _d}}}({G_{{\theta _g}}}({I^{LR}})))$
\end{center}

Existing super-resolution image reconstruction methods usually use MSE as the content loss function. However, MSE will make the reconstructed image too smooth and the resulting image is too artificial. In order to improve that, we introduce a new adaptive robust loss function\cite{Barron2019A}, which can independently learn the hyper-parameters set in the function while training the network, and it is able to reduce the time of searching the best hyper-parameters manually. The general form of the adaptive robust loss function is:

\begin{center}
$f(x,\alpha ,c) = \frac{{|\alpha  - 2|}}{\alpha }\left( {{{\left( {\frac{{{{\left( {{x \mathord{\left/
 {\vphantom {x c}} \right.
 \kern-\nulldelimiterspace} c}} \right)}^2}}}{{|\alpha  - 2|}} + 1} \right)}^{\frac{\alpha }{2}}} - 1} \right)$
\end{center}

$\alpha $ is a hyper-parameter that controls its robustness, and different values correspond to different loss functions. $c$ is a scale parameter.

In order to improve the quality of the reconstructed picture, the loss function of the generator also adds the perceptual loss. VGG is used to extract the high-level features of the picture, and the high-dimensional error between the real picture and the reconstructed high-resolution picture is calculated. The formula is as follows:

\begin{center}
$los{s_{VGG/i,j}} = \frac{1}{{{W_{i,j}}{H_{i,j}}}}\sum\limits_{x = 1}^{{W_{i,j}}} {\sum\limits_{y = 1}^{{H_{i,j}}} {{{({\Phi _{i,j}}{{({I^{HR}})}_{x,y}} - {\Phi _{i,j}}{{(G({I^{LR}}))}_{x,y}})}^2}} } $
\end{center}

$\Phi $ represents the feature map obtained through the VGG network.

In addition, our paper introduces the TVloss regularization into the generator objective function to constrain the difference between adjacent pixels in the picture. The objective function of the generator is:

\begin{center}
$los{s_G} = los{s_{adv}} + los{s_{f(\alpha ,x,c)}} + los{s_{VGG}} + los{s_{TV{\rm{loss}}}}$
\end{center}

$los{s_{adv}}$ is the generator adversarial loss, $los{s_{f(\alpha ,x,c)}}$ is the adaptive robust loss, $los{s_{VGG}}$ is the perceptual loss, and $los{s_{TV{\rm{loss}}}}$ is the regular loss.

\subsubsection{Discriminator structure}

The discriminator network draws on the discriminator network structure in SRGAN, as shown in Figure 1. In order to obtain more image feature information, we use 5*5 convolution kernels instead of 3*3 convolution kernels to expand the range of receptive fields. At the same time, we reduce the number of network layers. At the end of the network, a 1*1 convolution kernel is used, which is conducive to the trained model adapting to different sizes of test samples. The objective function of the discriminator network is:

\begin{center}
$\mathop {\max }\limits_D  = {E_{{I^{HR}} \sim {p_{data}}({I^{HR}})}}\log ({D_{{\theta _d}}}({I^{HR}})) + {E_{{I^{LR}} \sim {p_{data}}({I^{LR}})}}\log (1 - {D_{{\theta _d}}}({G_{{\theta _g}}}({I^{LR}})))$
\end{center}

The overall objective function of the network is:

\begin{center}
$\mathop {\min }\limits_G \mathop {\max }\limits_D V(D,G) = {E_{{I^{HR}} \sim {p_{data}}({I^{HR}})}}\log ({D_{{\theta _d}}}({I^{HR}})) + {E_{{I^{LR}} \sim {p_{data}}({I^{LR}})}}\log (1 - {D_{{\theta _d}}}({G_{{\theta _g}}}({I^{LR}})))$
\end{center}

Among them, the overall objective function not only includes the adversarial loss in GAN, but also includes the various constraints introduced above to jointly promote network training.

\section{Experiment}

This section first introduces the experimental conditions and training details, then introduces the data set and evaluation indicators, and finally shows the experimental results and analyzes them.

\subsection{Introduction to experimental conditions and training details}

The experiments are conducted on the Windows10 and NVIDIA 2080Ti server. The CUDA version is 10.2. All source programs are written in python language, implemented on the pytorch framework, pytorch version is 1.6. The discriminator uses a 5*5 convolution kernel to replace a 3*3 convolution kernel. The convolution kernel size of the generator is all 3*3. The optimized learning model method used for training is RMSprop(root mean square prop), which the parameter is 0.9. The $lr$(learning rate) is 0.0005. After reaching a certain number of iterations 20 epochs during the training process, $lr$ turns out to be 0.0001. The batch size is 64.

\subsection{Datasets introduction}

The model training and test datasets are natural images, and the VOC2012\cite{Ledig2016Photo} is used for model training and testing. First, preprocessing operations on the training set and test set in the VOC2012 set of the training network. The image are randomly clipped with a size of 128*128, and then down-sampled with a scaling factor of 2, 4 and 8. Datasets such as set5\cite{2012Low}, set14\cite{2010On}, and BSD100\cite{2011Contour} are used as performance tests. The picture are accessible directly from the trained model, while there are pictures in the set14 data set as grayscale images. The network is trained with 3 channels.

\subsection{Evaluation index}

Our paper evaluates the reconstructed super-resolution images under three evaluation indicators: PSNR, SSIM and MOS.

PSNR is often used to compare the differences between corresponding pixels. SSIM\cite{Zhou2004Image} is a measure based on brightness, contrast and structure, etc, to compare the similarity between corresponding image. MOS is used to evaluate the real visual observation of image. Above of methods verify the validity of the network structure.

The calculation of PSNR needs to calculate the mean square error MSE first, the overall formula is as follows:

\begin{center}
$MSE = \frac{1}{n}\sum\limits_{i = 1}^n {{{({I^{HR}} - {I^{SR}})}^2}} $
\end{center}

\begin{center}
$PSNR = 10 \cdot {\log _{10}}\left( {{1 \mathord{\left/
 {\vphantom {1 {MSE}}} \right.
 \kern-\nulldelimiterspace} {MSE}}} \right)$
\end{center}

The calculation formula of SSIM is:

\begin{center}
$SSIM = \frac{{\left( {2{\mu _{{I^{SR}}}}{\mu _{{I^{HR}}}} + {C_1}} \right)\left( {2{\sigma _{{I^{SR}}{I^{HR}}}} + {C_2}} \right)}}{{\left( {\mu _{{I^{SR}}}^2 + \mu _{{I^{HR}}}^2 + C1} \right)\left( {\sigma _{{I^{SR}}}^2 + \sigma _{{I^{HR}}}^2 + C1} \right)}}$
\end{center}

Where ${\mu _{{I^{SR}}}}$ is the mean of ${I^{SR}}$, ${\mu _{{I^{HR}}}}$ is the mean of ${I^{HR}}$, ${\sigma _{{I^{SR}}{I^{HR}}}}$ is  the covariance of ${I^{SR}}$ and ${I^{HR}}$,  and $\sigma _{{I^{SR}}}^2$ and $\sigma _{{I^{HR}}}^2$ is the variance of ${I^{SR}}$ and ${I^{HR}}$.

SRGAN\cite{Ledig2016Photo} introduces the MOS evaluation method into the super-resolution image reconstruction task. MOS is a subjective evaluation method that contains 5 levels. We collect the scores of the reconstructed image from a user study, with 1-2 denotes as poor, 2-3 points are average, 3-4 points are good, 4-5 points are excellent. In our paper, 24 volunteers scored the reconstructed super-resolution images, and then calculate the average result of the score.

The above evaluation methods are that the larger the value, the smaller the gap between the reconstructed image and the real image, and the higher the quality of the reconstructed image.

\subsection{Experimental results and analysis}

This section mainly shows the results of super-resolution image reconstruction and evaluation index results of different algorithms, and the analysis of the results.

The following is a brief description of the reconstruction algorithms that need to be compared:

\begin{itemize}

\item A reconstruction algorithm based on Bicubic, which is cubic linear interpolation of low-resolution images.

\item Based on the algorithm of SRCNN\cite{Chao2014Learning}. In this paper, CNN is used for super-resolution image reconstruction for the first time. It is a shallow network. The reconstruction process is divided into three stages, making a great contribution to the neural network processing super-resolution images.

\item The algorithm based on FSRCNN\cite{Chao2016Accelerating} is also a kind of shallow neural network, which improves the mapping layer and activation function of SRCNN, and uses the deconvolution layer at the end of the network to speed up the training speed. It is better than SRCNN.

\item The algorithm based on MSRN\cite{Li2018Multi} is a deep network that uses different sizes of convolution kernels to extract features and fuse the extracted features to obtain more detailed information. Although good results are obtained, the network is more complicated.

\item Based on the algorithm of SRGAN\cite{Ledig2016Photo} and the GAN structure, the min-max game method is used to train the network. The generator uses the residual network to learn the details of the image. At the same time, a perceptual loss function is proposed, which increases the content loss on the basis of the adversarial loss. Content loss helps guide the reconstruction of super-resolution images with clear details.

\item Based on the algorithm of multi-scale generative adversarial network\cite{Liu2020Image}, it improves SRGAN. The generator uses a residual network with two sub-structures, and performs information fusion to learn more and more detailed information of the image.

\item In the method of our paper, the generator uses a self-calibrated convolutional network with 3*3 size convolution kernel. Without increasing network parameters, the network structure is simple. The network structure is divided into different parts and each part has different functions. Due to feature fusion, richer detailed information can be extracted and used. The reconstructed super-resolution image has a better effect.

\end{itemize}

The reconstruction results of the above various algorithms are shown in Figure 4. It can be seen from the figure that the algorithm in our paper has a good visual effect and a high similarity to the real picture.

\begin{figure}[t]
\centering
\includegraphics[scale=0.6]{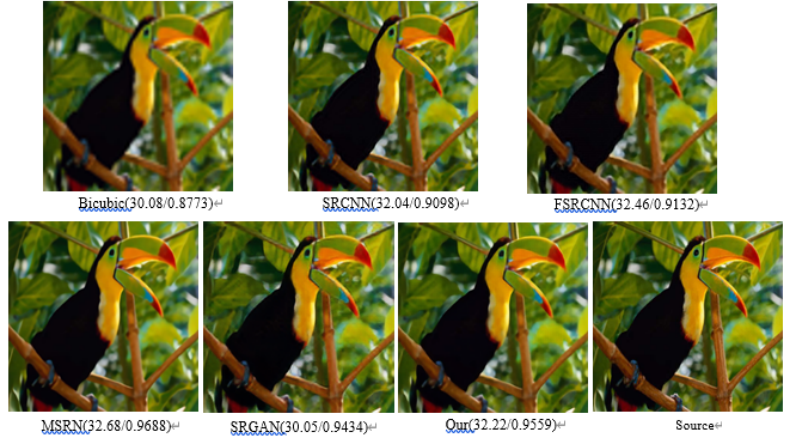}
\caption{Visualization of the result}
\end{figure}

Table 1-3 shows the experimental results of different algorithms in different datasets, with using different evaluation indicators when the scaling factor is 4.
\begin{table*}[htb]
\centering
\caption{Quantitative evaluation results in set5} \label{t1}
\begin{tabular}{ccccccccc}
\hline
Method&Scale&Bicubic&SRCNN&FSRCNN&MSRN&SRGAN&Multi-GAN&Our\\
\hline
PSNR& $ \times $4&28.47&30.25&30.50&32.07&28.68&29.64&30.14\\
SSIM& $ \times $4&0.8184&0.8647&0.8644&0.8903&0.9202&0.8557&0.9304\\
MOS& $ \times $4&1.96&2.95&3.23&4.04&3.69&3.62&3.89\\
\hline
\end{tabular}
\end{table*}

\begin{table*}[htb]
\centering
\caption{Quantitative evaluation results in set14} \label{t1}
\begin{tabular}{ccccccccc}
\hline
Method&Scale&Bicubic&SRCNN&FSRCNN&MSRN&SRGAN&Multi-GAN&Our\\
\hline
PSNR& $ \times $4&26.01&27.13&27.29&28.60&25.85&26.73&26.72\\
SSIM& $ \times $4&0.7250&0.7571&0.7578&0.7751&0.8451&0.7346&0.8546\\
MOS& $ \times $4&2.14&3.13&3.31&4.23&3.82&3.69&4.04\\
\hline
\end{tabular}
\end{table*}
\begin{table*}[htb]
\centering
\caption{Quantitative evaluation results in BSD100} \label{t1}
\begin{tabular}{ccccccccc}
\hline
Method&Scale&Bicubic&SRCNN&FSRCNN&MSRN&SRGAN&Multi-GAN&Our\\
\hline
PSNR& $ \times $4&26.02&26.85&26.94&27.52&25.69&25.24&26.02\\
SSIM& $ \times $4&0.6810&0.7192&0.7200&0.7273&0.8033&0.6634&0.8116\\
MOS& $ \times $4&2.08&2.96&3.15&3.96&3.62&3.58&3.82\\
\hline
\end{tabular}
\end{table*}
Comparing the results from Table 1-3, the algorithm in our paper has good results on the evaluation index SSIM. Under the MOS evaluation index, the method in our paper is superior to other algorithms, except for the MSNR algorithm, which proves that the algorithm has certain advantages. The algorithm in our paper is proposed based on the GAN framework. SRGAN and MutliGAN in the table are also based on the GAN framework, and all of them use VOC2012 as training datasets, while other algorithms use DIV$2$K \cite{Agustsson2017NTIRE} as training datasets. VOC2012 data set does not contain the image samples in the test set, and even samples of the same type. Therefore, the sample information cannot be learned, and the average result under the PSNR indicator is poor. The algorithm in our paper are improved that compared with the Mutli-GAN based on the SRGAN structure. Under the three evaluation indicators, our algorithm in three algorithms base on GAN framework has the best results. The experimental results of each comparison algorithm are slightly fluctuating compared with the original text, but they are all relatively close to the results in the text, (except for MSRN and Multi-GAN using the experimental data, the result of other algorithm is run on ours own machine).
\subsection{Loss function comparison experiment}

Our paper introduces an adaptive robust loss function in the adversarial loss. In order to prove the effectiveness of replacing the content loss function MSE with an adaptive robust loss function, we compare the evaluation results of the network on different performance test data sets before and after the replacement. The experimental results are shown in table 4.

\begin{table*}[htb]
\centering
\caption{Quantitative evaluation results of comparative experiments} \label{t1}
\begin{tabular}{ccccc}
\hline
\multirow{2}{*}{Loss}& \multirow{2}{*}{Scale} & set5 & set14 & BSD100\\
&   & PSNR/SSIM & PSNR/SSIM & PSNR/SSIM\\
\hline
Adaptive robust loss&$ \times $4&30.14/0.9304&26.72/0.8546&26.02/0.8116\\
MSE&$ \times $4&29.31/0.9193&26.14/0.8446&25.74/0.8056\\
\hline
\end{tabular}
\end{table*}

From table 4, it can be seen that the adaptive robust loss function can effectively improve the quality of the reconstructed image, and the performance is better than the MSE loss function under different evaluation indicators. Meanwhile, it alleviates the over-smoothing reconstruction of the image caused by the MSE loss function.

\section{Conclusion}

In this paper, we propose a super-resolution image reconstruction method based on a generative adversarial network, and gives a low-resolution image that can reconstruct the super-resolution image corresponding to the low-resolution image. The generator of the generative adversarial network uses a self-calibrated convolutional network can extract the rich detailed information of low-resolution images and make full use of it. It not only expands the feature space scale range, but also maintains the dependency between the extracted feature channels. In addition, the adversarial loss includes adaptive robust loss to help the network stably train, and at the same time alleviates the problem of over-smoothing reconstructed image caused by the MSE loss function. Finally, experiments show that the method in our paper has better human eye observation effects than other algorithms. Compared with other various algorithms, it is superior under the SSIM evaluation index. But there is no significant improvement under the PSNR evaluation index, and further research is needed.

\section*{Acknowledgment}
This work are sponsored by the key project of the National Natural Science Foundation of China (61602421), China Postdoctoral Foundation (2016M600584), Aviation Science Foundation (201828X4001) and Zhengzhou University.

\bibliographystyle{IEEEtran}
\bibliography{reference}
\ifCLASSOPTIONcaptionsoff
  \newpage
\fi

\end{document}